\documentclass[10pt,numbers]{elsarticle}
\makeatletter
\def\ps@pprintTitle{%
 \let\@oddhead\@empty
 \let\@evenhead\@empty
 \def\@oddfoot{\centerline{\thepage}}%
 \let\@evenfoot\@oddfoot}
\makeatother

\usepackage{amssymb,amsmath,mathtools}
\usepackage{cases,resizegather}
\usepackage[left=35mm,right=35mm]{geometry}
\usepackage{graphicx}

\newtheorem{thm}{Theorem}
\newtheorem{lem}[thm]{Lemma}
\newtheorem{cor}[thm]{Corollary}

\newtheorem{rmk}{Remark}

\newdefinition{defi}{Definition}
\newproof{prf}{Proof}

\newcommand{\GF}[1]{\mathbb{F}_{2^{#1}}}
\newcommand{\Tr}{\mathrm{Tr}}
\newcommand*{\QED}{\hfill\ensuremath{\square}}
\newcommand{\Hb}{\mathsf{H}}
\newcommand{\Vb}{\mathsf{V}}
\newcommand{\Gal}{\mathrm{Gal}}
\renewcommand{\eqref}[1]{Eq.(\ref{#1})}

\begin{document}

\begin{frontmatter}



\title{Further Results of the Cryptographic Properties on the Butterfly Structures}


\author[amss]{Shihui Fu\corref{corauth}}
\ead{fushihui@amss.ac.cn}
\author[amss]{Xiutao Feng}
\ead{fengxt@amss.ac.cn}

\address[amss]{Key Laboratory of Mathematics Mechanization, Academy of Mathematics and Systems Science, Chinese Academy of Sciences, Beijing 100190, China}
\cortext[corauth]{Corresponding author}

\begin{abstract}
Recently, a new structure called butterfly introduced by Perrin et at. is attractive for that it has very good cryptographic properties: the differential uniformity is at most equal to 4 and algebraic degree is also very high when exponent $e=3$. It is conjecture that the nonlinearity is also optimal for every odd $k$, which was proposed as a open problem.\par

\noindent In this paper, we further study the butterfly structures and show that these structure with exponent $e=2^i+1$ have also very good cryptographic properties. More importantly, we prove in theory the nonlinearity is optimal for every odd $k$, which completely solve the open problem.\par

\noindent Finally, we study the butter structures with trivial coefficient and show these butterflies have also optimal nonlinearity. Furthermore, we show that the closed butterflies with trivial coefficient are bijective as well, which also can be used to serve as a cryptographic primitive.\par
\end{abstract}

\begin{keyword}


S-box \sep APN \sep butterfly structure \sep permutation \sep differential uniformity \sep nonlinearity
\MSC[\hspace{-1.3mm}] 94A60 \sep 11T71 \sep 14G50
\end{keyword}

\end{frontmatter}



\section{Introduction}
Many block ciphers use substitution boxes (S-boxes) to serve as the confusion part. To obtain a correct decryption, S-boxes are usually chosen to be permutation over a finite field with characteristic 2 and even extension degree, i.e., $\GF{2k}$. For ease of implementation and to have good cryptographic properties to resist various kinds of cryptographic attacks, S-boxes used in block ciphers should possess low differential uniformity to resistant differential attack \cite{DiffUni}, high nonlinearities to resistant linear attack \cite{NonLin}.\par

It is well known that for any function defined over $\GF{n}$, the lowest differential uniformity is 2, and the functions achieving this value are called almost perfect nonlinear (APN) functions. Unfortunately, it is very hard to construct APN permutations for $n$ even. Up to now, only one APN permutation over $\GF{6}$ has been found. To find any other APN permutations over $\GF{n}$ for even $n$ is called the the BIG APN problem \cite{APNPeru}. \par

Therefore, a natural tradeoff method is to use differentially 4-uniform permutations as S-boxes. For instance, the AES (advanced encryption standard) uses a differentially 4 uniform function, namely, the inverse function as S-boxes. Hence to provide more choices for the S-boxes, it is of significant importance to construction more infinite classes differentially 4-uniform permutations over $\GF{2m}$ with good cryptographic properties.\par

Recently in \cite{Butterfly}, Perrin et al. introduced a new structure called butterfly structure and showed that these structures with exponent $e=3\times 2^t$ always have differential uniformity at most 4 when $n=2k$ with $k$ odd. The authors also verified experimentally that the nonlinearity of the butterfly structure is equal to $2^{2k-1}-2^k$ for $k=3,5,7$. However, they could not prove it in the general case and conjecture that equality is true for every odd $k$.\par

In \cite{Butterfly}, Li and Wang proposed a construction with 3-round Feistel structure, which is actually a particular cases of the butterfly structure with coefficient 1. They proved that this structure have differential uniformity 4 and algebraic degree $k$.\par

In general, the cost of hardware implementation of nonlinear functions is increasing with its input and output size. Thus implementing functions over subfield often cost much less than implementing functions over the larger field. It is an huge advantage of constructing S-boxes over $\GF{k}^2$ with butterfly structure for that we only need to implement the exponent functions over $\GF{k}$. Therefore, comparing with $2k$-bit S-boxes constructed directly with permutation over $\GF{2k}$, S-boxes over $\GF{k}^2$ constructed via butterfly structure with exponent function over $\GF{k}$ cost much less in hardware implementation. \par

In the present paper, we further revisit the butterfly structure and prove that these structures with exponent $e=(2^i+1)\times 2^t$ also have differential uniformity at most 4 when $n=2k$ with $k$ odd. Moveover, we prove theoretically that the nonlinearity equality is true for every odd $k$, which means these constructions have the optimal nonlinearity in the sense that no known functions of a field of even size have a higher nonlinearity. Finally, we also study the butterfly structure with trivial coefficient $\alpha=1$, and show that nonlinearity are also optimal. Besides, the closed butterfly structure with trivial coefficient are also a permutation, which also can be used to serve as a cryptographic primitive. \par

The rest of this paper is organized as follows. In the next section, we recall needed knowledge and some necessary definitions and results. In Section \ref{sec:nontr}, we show that the differential uniformity of butterfly structures with branch size $k$ odd, exponent $e=(2^i+1)\times 2^t$ and nontrivial coefficient are at most equal to 4, and the nonlinearity are optimal as well. In section \ref{sec:trivl}, we further study the butterfly structure with coefficient $\alpha=1$ and show that the structures have also the optimal nonlinearity. The proof of bijective of closed butterfly is also given in this section. The concluding remarks are given in Section \ref{sec:concl}. \par

\section{Preliminaries}\label{sec:preli}

Throughout this paper, let $n$ is a positive integer, $\mathbb{F}_{2^n}$ be the finite field with $2^n$ elements and $\mathbb{F}_{2^n}^*$ be the multiplicative group of order $2^n-1$.
A function $F:\GF{n}\rightarrow \GF{n}$ can be represented uniquely in a polynomial form in $\GF{n}[x]/\langle x^{2^n}+x\rangle$ as
\[
F(x)=\sum_{i=0}^{2^n-1}c_ix^i,\quad c_i\in\GF{n}.
\]
For any $l$, $0\leq l \leq 2^n-1$, the number $w_2(l)$ of the nonzero coefficients $l_j\in\mathbb{F}_2$ in the binary expansion $l=\sum_{j=0}^{n-1}l_j2^j$ is called the 2-weight of $l$. The algebraic degree of $F$, denoted by $\deg(F)$, is equal to the maximum 2-weight of $i$ such that $c_i\neq 0$.
\begin{defi}[\cite{DefDiff}]
For a function $F:\GF{n}\rightarrow \GF{n}$, the differential uniformity of $F(x)$ is denoted as
\[
\delta_F = \max\{\delta_F(a,b):a\in\mathbb{F}_{2^n}^{*},b\in\mathbb{F}_{2^n}\},
\]
where $\delta_F(a,b) = |\{x\in\mathbb{F}_{2^n}:F(x+a)+F(x)=b\}|$. The differential spectrum of $F(x)$ is the set
\[
\{\delta_F(a,b):a\in\mathbb{F}_{2^n}^{*},b\in\mathbb{F}_{2^n}\}.
\]
\end{defi}\par

$F(x)$ is called differential $\delta$-uniform if $\delta_F = \delta$. It is easy to see that if $x_0$ is a solution of $F(x+a)+F(x)=b$, so does $x_0+a$. Thus the lower bound on differential uniformity of $F(x)$ is 2. The functions which achieve this bound are called almost perfect nonlinear functions (APN).\par

For any function $F:\mathbb{F}_{2^n}\rightarrow\mathbb{F}_{2^n}$, we define the Walsh transform of $F$ as
\[
\mathcal{W}_F(a,b)=\sum_{x\in\GF{n}}(-1)^{\Tr (bF(x)+ax)},\quad a,b\in\GF{n},
\]
where $\Tr(x)=x+x^2+\cdots +x^{2^{n-1}}$ is the absolute trace function from $\GF{n}$ to $\mathbb{F}_2$. The set $\Lambda_F=\{\mathcal{W}_F(a,b):a\in\GF{n},b\in\GF{n}^*\}$ is called the Walsh spectrum of the function $F$.\par

The nonlinearity of $F$ is defined as
\[
\mathcal{NL}(F)=2^{n-1}-\frac{1}{2}\max_{a\in\GF{n},b\in\GF{n}^*}|\mathcal{W}_F(a,b)|.
\]
It is known that if $n$ is odd the nonlinearity of $F$ satisfies the inequality $\mathcal{NL}(F)\leq 2^{n-1}-2^{\frac{n-1}{2}}$ \cite{BoundOdd} and in case of equality $F$ is called almost bent (AB). The notion of AB function is closely connected the notion of APN function. AB function exist only for $n$ odd and oppose an optimum resistance to linear cryptanalysis. Besides, every AB function is APN, and in the $n$ odd case, any quadratic APN function is AB function. A comprehensive survey on APN and AB functions can be found in \cite{ABandAPN,AlgDegree}.\par

While $n$ is even, the upper bound of nonlinearity is still open. The known maximum nonlinearity is $2^{n-1}-2^{\frac{n}{2}}$ \cite{BoundEven}. It is conjectured that $\mathcal{NL}(F)$ is upper bounded by $2^{n-1}-2^{\frac{n}{2}}$. These functions which meet this bound are usually called optimal (maximal) nonlinear functions.\par

For two functions $F,G:\GF{n}\rightarrow \GF{n}$ are called extended affine equivalent (EA-equivalent), if $G(x)=A_1(F(A_2(x)))+A_3(x)$, where $A_1(x)$ and $A_2(x)$ are affine permutations on $\GF{n}$ and $A_3(x)$ is an affine function over $\GF{n}$. They are called CCZ-equivalent (Carlet-Charpin-Zinoviev equivalent) if there is an affine permutation over $\GF{n}^2$ which maps $\mathcal{G}_F$ to $\mathcal{G}_G$ , where $\mathcal{G}_F = \{(x,F(x)):x\in \GF{n}\}$ is the graph of $F$, and $\mathcal{G}_G$ is the graph of $G$.\par

It is well known that EA-equivalence implies CCZ-equivalence, but not vice versa. Differential uniformity, nonlinearity and Walsh spectrum are invariants of both EA-equivalence and CCZ-equivalence. Algebraic degree is preserved by EA-equivalence, but not CCZ-equivalence. However, neither EA-equivalence nor CCZ-equivalence preserve permutations.\par

\begin{defi}[\cite{Butterfly}]
  Let $k$ be an integer and $\alpha\in\GF{k}$, $e$ be an integer such that $x\mapsto x^e$ is a permutation over $\GF{k}$ and $R_z[e,\alpha](x)=(x+\alpha z)^e+z^e$ be the keyed permutation. The Butterfly Structures are defined as follows:\\
  (a) the Open Butterfly Structure with branch size $k$, exponent $e$ and coefficient $\alpha$ is the function denoted $\Hb_e^{\alpha}$ defined by:
  \[
  \Hb_e^{\alpha}(x,y)=(R_{R_y^{-1}[e,\alpha](x)}(y),R_y^{-1}[e,\alpha](x)),
  \]
  (b) the Closed Butterfly Structure with branch size $k$, exponent $e$ and coefficient $\alpha$ is the function denoted $\Vb_e^{\alpha}$ defined by:
  \[
  \Vb_e^{\alpha}(x,y)=(R_x[e,\alpha](y),R_y[e,\alpha](x)).
  \]
\end{defi}\par

Note that $\Hb_e^{\alpha}$ always a permutation over $\GF{k}^2$, while $\Vb_e^{\alpha}$ maybe not. In fact, $\Hb_e^{\alpha}$ is an involution over $\GF{k}^2$, which means $\Hb_e^{\alpha}(\Hb_e^{\alpha}(x,y))=(x,y)$. Furthermore, the permutation $\Hb_e^{\alpha}$ and the function $\Vb_e^{\alpha}$ are CCZ-equivalent \cite{Butterfly}.\par

The Walsh spectrum and algebraic degree of a function $F(x)\in\GF{n}[x]$ have the following containment relationships, which is needed to proof our results.
\begin{lem}[\cite{AlgDegree}]\label{lem:aldeg}
  Suppose $F(x)\in\GF{n}[x]$. If $2^l\big|\mathcal{W}_F(a,b)$ for any $a,b\in\GF{k}$ with $b\neq 0$, then the algebraic degree of $F(x)$ is at most equal to $n-l+1$.
\end{lem}

Let $L$ be a extension of field $K$, $\sigma\in\Gal(L/K)$, and polynomial $w(t)=\sum_{j=0}^{l}c_jt^j\in L[t]$. Then $w(t)$ acts on a element $x$ of $L$ is defined as $w(\sigma)x=\sum_{j=0}^{l}c_j\sigma^j(x)$. The following lemma is also needed.\par

\begin{lem}[\cite{BilForms}]
Let $L$ be a cyclic Galois extension of $K$ of degree $n$ and suppose that $\sigma$ generates the Galois group of $L$ over $K$. Let $m$ be an integer satisfying $1\leq m\leq n$ and let $w(t)$ be a polynomial of degree $m$ in $L[t]$. Let
\[
  R=\{x\in L : w(\sigma)x=0\}.
\]
Then we have $\dim_K R\leq m$.
\end{lem}

It is well known that Frobenius automorphism $\sigma(x)=x^2$ generates the cyclic group $\Gal(\GF{k}/\GF{})\simeq\mathbb{Z}/k\mathbb{Z}$. Moveover, if $\gcd(i,k)=1$, then $\sigma^i(x)=x^{2^i}$ is also a generators. We have the following corollary.
\begin{cor}\label{cor:4sol2}
  Suppose $k$ is an integer and $\gcd(i,k)=1$. For any $c_1,c_2,c_3\in\GF{k}$ with not all zero, then the following equation
  \[
    c_1x^{2^{2i}}+c_2x^{2^i}+c_3x=0
  \]
  has at most 4 solutions over $\GF{k}$.
\end{cor}

Moveover, if $k$ is an odd integer and $\gcd(i,k)=1$, then $\gcd(2i,k)=1$. The next corollary is obviously.
\begin{cor}\label{cor:4sol4}
  Suppose $k$ is an odd integer and $\gcd(i,k)=1$. For any $c_1,c_2,c_3\in\GF{k}$ with not all zero, then the following equation
  \[
    c_1x^{2^{4i}}+c_2x^{2^{2i}}+c_3x=0
  \]
  has at most 4 solutions over $\GF{k}$.
\end{cor}

\section{Butterfly Structure with $\alpha\neq 0,1$}\label{sec:nontr}

When $k$ is odd, $\gcd(i,k)=1$, we have also $\gcd(2i,k)=1$, which implies that $\gcd(2^i\pm 1, 2^k-1)=1$. Both maps $x\mapsto x^{2^i+1}$ and $x\mapsto x^{2^i-1}$ are bijective over $\GF{k}$.\par

In this section, we study the butterfly structures with $\alpha\neq 0,1$ for block sizes $2k$ ($k$ odd). In section \ref{sbs:dfuni}, we show that these structures are always differential 4-uniform. In section \ref{sbs:aldeg}, the algebraic degree is given. In section \ref{sbs:nonli}, we show that the nonlinearity of these structures are optimal.\par

\subsection{Differential uniformity}\label{sbs:dfuni}

In order to characterize the differential uniformity, we need the following lemma firstly.\par

\begin{lem}\label{lem:dfudp}
  Let $k$ is an odd integer and $\gcd(i,k)=1$. Then for any $\alpha\in\GF{k}$ with $\alpha\neq 0,1 $, the following system of equations in variables $u,v$
  \begin{numcases}{}
     (\alpha v+u)\left(\alpha(\alpha u+v)^{2^i}+u^{2^i}\right)+(\alpha v+u)^{2^i}\left(\alpha^{2^i}(\alpha u+v)+u\right)=0, \label{equ:auv01}\\
     (\alpha v+u)(\alpha u+v)+\left(\alpha^{2^i}(\alpha u+v)+u\right)\left(\alpha^{2^i}(\alpha v+u)+v\right)=0, \label{equ:auv02}\\
     (\alpha v+u)(\alpha u+v)^{2^i}+\left(\alpha^{2^i}(\alpha u+v)+u\right)\left(\alpha(\alpha v+u)^{2^i}+v^{2^i}\right)=0 \label{equ:auv03}
  \end{numcases}
  holds over $\GF{k}$ if and only if $u,v$ satisfying $\alpha v+u=0$ and $\alpha^{2^i}(\alpha u+v)+u=0$.
\end{lem}

\begin{prf}
  The sufficiency is obvious. Now we show the necessary.\par

  If $\alpha v+u=0, \alpha^{2^i}(\alpha u+v)+u\neq 0$, then $u\neq 0, v\neq 0$. \eqref{equ:auv02} becomes $\left(\alpha^{2^i}(\alpha u+v)+u\right)v=0$, which contradicts that $\alpha^{2^i}(\alpha u+v)+u\neq 0$ and $v\neq 0$.\par

  If $\alpha v+u\neq0, \alpha^{2^i}(\alpha u+v)+u=0$, then from \eqref{equ:auv02} we get $\alpha u+v=0$. Note that $\alpha^{2^i}(\alpha u+v)+u=0$, we have $u=v=0$, which contradicts that $\alpha v+u\neq0$.\par

  Now we always assume that $\alpha v+u\neq0, \alpha^{2^i}(\alpha u+v)+u\neq 0$.\par

  If $u=0$, then $v\neq 0$. According to \eqref{equ:auv01}, We obtain $(\alpha^{2^i}+\alpha)^2=0$, which is impossible since $\alpha\neq 0,1$ and $\gcd(i,k)=1$.\par

  If $v=0$, then $u\neq 0$. From \eqref{equ:auv02} and \eqref{equ:auv03}, we have $\alpha^{2^{i+1}+1}+\alpha^{2^i}+\alpha=0$ and $\alpha^{2^i+2}+\alpha^{2^i}+\alpha=0$. Hence $\alpha^{2^{i+1}+1}+\alpha^{2^i+2}=\alpha^{2^i+1}(\alpha^{2^i}+\alpha)=0$, which is impossible as well.\par

  We also assume that $u\neq 0, v\neq 0$.\par

  Then $\alpha(\alpha u+v)^{2^i}+u^{2^i}\neq 0$, otherwise, from \eqref{equ:auv01} we have $\alpha v+u=0$ or $\alpha^{2^i}(\alpha u+v)+u=0$, which contradicts the assumption. And also $\alpha u+v\neq 0$, otherwise, from \eqref{equ:auv03} we have $u=0$ or $\alpha^{2^i}(\alpha u+v)+u=0$, which also contradicts the assumption.\par

  According to \eqref{equ:auv01} and \eqref{equ:auv03}, we have
  \begin{equation}\label{equ:auv13}
    \frac{(\alpha v+u)^{2^i}}{\alpha(\alpha u+v)^{2^i}+u^{2^i}}=\frac{\alpha v+u}{\alpha^{2^i}(\alpha u+v)+u}=\frac{\alpha(\alpha v+u)^{2^i}+v^{2^i}}{(\alpha u+v)^{2^i}}
  \end{equation}
  To simplify expression, we denote $\alpha=\beta^{2^i}$, then $\beta\neq 0,1$. From \eqref{equ:auv13} we get
  \[
  \frac{(\alpha v+u)^{2^i}}{(\beta(\alpha u+v)+u)^{2^i}}=\frac{(\beta(\alpha v+u)+v)^{2^i}}{(\alpha u+v)^{2^i}},
  \]
  which is equivalent to
  \begin{equation}\label{equ:auv04}
    (\alpha v+u)(\alpha u+v)=(\beta(\alpha u+v)+u)(\beta(\alpha v+u)+v).
  \end{equation}
  We combine the above equation with \eqref{equ:auv02} and get
  \[
  (\beta(\alpha u+v)+u)(\beta(\alpha v+u)+v)=\left(\alpha^{2^i}(\alpha u+v)+u\right)\left(\alpha^{2^i}(\alpha v+u)+v\right),
  \]
  which can be simplify to that
  \[
  (\alpha^{2^i}+\beta)^2(\alpha u+v)(\alpha v+u)+(\alpha^{2^i}+\beta)v(\alpha u+v)+(\alpha^{2^i}+\beta)u(\alpha v+u)=0.
  \]
  Notice that $\alpha^{2^i}+\beta=\beta^{2^{2i}}+\beta\neq 0$ since $\beta\neq 0,1$, $k$ odd and $\gcd(i,k)=1$, dividing by $\alpha^{2^i}+\beta$ we get
  \[
  (\alpha^{2^i}+\beta)(\alpha u+v)(\alpha v+u)+v(\alpha u+v)+u(\alpha v+u)=0.
  \]
  Denote $u=\lambda v$, then $\lambda\neq 0$, the above equation becomes
  \begin{equation}\label{equ:auv05}
    (\alpha^{2^i+1}+\alpha\beta+1)\lambda^2+(\alpha^{2^i}+\beta)(\alpha^2+1)\lambda+(\alpha^{2^i+1}+\alpha\beta+1)=0.
  \end{equation}
  In case of $\alpha^{2^i+1}+\alpha\beta+1=0$. Note that $(\alpha^{2^i}+\beta)(\alpha^2+1)\neq 0$, then $\lambda=0$, which is a contradiction. Hence $\alpha^{2^i+1}+\alpha\beta+1\neq 0$. By dividing $\alpha^{2^i+1}+\alpha\beta+1$, \eqref{equ:auv05} is equivalent to
  \begin{equation}\label{equ:auv06}
    \lambda^2+\frac{(\alpha^{2^i}+\beta)(\alpha^2+1)}{\alpha^{2^i+1}+\alpha\beta+1}\lambda+1=0
  \end{equation}\par

  We replace \eqref{equ:auv02} by $u=\lambda v$ and get
  \[
    (\alpha^{2^{i+1}+1}+\alpha^{2^i}+\alpha)\lambda^2+(\alpha^{2^{i+1}+2}+\alpha^{2^{i+1}}+\alpha^2)\lambda+(\alpha^{2^{i+1}+1}+\alpha^{2^i}+\alpha)=0
  \]
  Similarly, if $\alpha^{2^{i+1}+1}+\alpha^{2^i}+\alpha=0$, since $\alpha^{2^{i+1}+2}+\alpha^{2^{i+1}}+\alpha^2=(\alpha+1)^{2(2^i+1)}+1\neq 0$, we obtain $\lambda=0$, which is a contradiction. By dividing $\alpha^{2^{i+1}+1}+\alpha^{2^i}+\alpha$, the above equation becomes
  \begin{equation}\label{equ:auv07}
    \lambda^2+\frac{\alpha^{2^{i+1}+2}+\alpha^{2^{i+1}}+\alpha^2}{\alpha^{2^{i+1}+1}+\alpha^{2^i}+\alpha}\lambda+1=0.
  \end{equation}\par

  For each equation in variable $\lambda$ of \eqref{equ:auv06} and \eqref{equ:auv07}, either has two solutions or has no solution over $\GF{k}$. Moreover, the two solutions are the inverse of each other. It is readily to verify that the intersection of solutions sets of two equations is identical or empty. If the intersection is empty, then at least one of \eqref{equ:auv01}-(\ref{equ:auv03}) does not hold. If the two solutions sets is identical, we must have
  \[
  \frac{(\alpha^{2^i}+\beta)(\alpha^2+1)}{\alpha^{2^i+1}+\alpha\beta+1}=\frac{\alpha^{2^{i+1}+2}+\alpha^{2^{i+1}}+\alpha^2}{\alpha^{2^{i+1}+1}+\alpha^{2^i}+\alpha}.
  \]
  Replace $\alpha=\beta^{2^i}$, we deduce that
  \[
  \beta^{2^{2i}+2^{i+1}+1}+\beta^{2^{2i}+1}+\beta^{2^{2i}+2^i}+\beta^{2^{i+1}}+\beta^{2^i+1}=0,
  \]
  which, furthermore, is equal to the following equation
  \[
  \beta^{(2^i+1)^2}=(\beta^{2^i}+\beta)^{2^i+1}.
  \]
  Notice that $x^{2^i+1}$ is a permutation over $\GF{k}$, hence we have
  \[
  \beta^{2^i+1}+\beta^{2^i}+\beta=(\beta+1)^{2^i+1}+1=0
  \]
  which is contradict to that $\beta\neq 0$. Hence, at least one of the \eqref{equ:auv01}-(\ref{equ:auv03}) does not hold.\par

  We complete the proof.\QED
\end{prf}

\begin{thm}
  Let $n$ is an odd integer and $\gcd(i,k)=1$. For any $0\leq t\leq k-1$, $\alpha\in\GF{k}$, and $\alpha\neq 0,1$, let $\Hb_{e}^{\alpha}$ and $\Vb_{e}^{\alpha}$ be the open and closed $2k$-bit butterflies structure with exponent $e=(2^i+1)\times 2^t$ and coefficient $\alpha$. Then the differential uniformity of both $\Hb_{e}^{\alpha}$ and $\Vb_{e}^{\alpha}$ is at most equal to 4.
\end{thm}

\begin{prf}
  As $\Hb_{e}^{\alpha}$ and $\Vb_{e}^{\alpha}$ are CCZ-equivalent, they have the same differential uniformity. It is sufficient to prove that the differential uniformity of $\Vb_{e}^{\alpha}$ is 4. Besides, the functions $\Vb_{e}^{\alpha}$ with exponent $e=(2^i+1)\times 2^t$ is affine equivalent to functions $\Vb_{e}^{\alpha}$ with exponent $e=2^i+1$. Thus it is sufficient to consider the case where the exponent is equal to $e=2^i+1$.\par

  Let $u,v,a,b\in\GF{k}$ and $(u,v)\neq(0,0)$. Then we need to prove that
  \begin{equation}\label{equ:odfuv}
    \Vb_{e}^{\alpha}(x,y)+\Vb_{e}^{\alpha}(x+u,y+v)=(a,b),
  \end{equation}
  namely, the following system of equations
  \begin{equation*}
    \begin{cases}
      \begin{array}{@{}l@{}}
      \left(\alpha^{2^i}(\alpha u+v)+u\right)x^{2^i}+\left(\alpha(\alpha u+v)^{2^i}+u^{2^i}\right)x\\
      \multicolumn{1}{r}{+(\alpha u+v)y^{2^i}+(\alpha u+v)^{2^i}y=(\alpha u+v)^{2^i+1}+u^{2^i+1}+a,}\\
      (\alpha v+u)x^{2^i}+(\alpha v+u)^{2^i}x\\
      \multicolumn{1}{r}{+\left(\alpha^{2^i}(\alpha v+u)+v\right)y^{2^i}+\left(\alpha(\alpha v+u)^{2^i}+v^{2^i}\right)y=(\alpha v+u)^{2^i+1}+v^{2^i+1}+b.}
    \end{array}
    \end{cases}
  \end{equation*}
  has at most 4 solutions over $\GF{k}^2$, which is equivalent to the linear homogeneous system of above equations
  \begin{numcases}{}
    \left(\alpha^{2^i}(\alpha u+v)+u\right)x^{2^i}+\left(\alpha(\alpha u+v)^{2^i}+u^{2^i}\right)x+(\alpha u+v)y^{2^i}+(\alpha u+v)^{2^i}y=0, \label{equ:hmdf1}\\
    (\alpha v+u)x^{2^i}+(\alpha v+u)^{2^i}x+\left(\alpha^{2^i}(\alpha v+u)+v\right)y^{2^i}+\left(\alpha(\alpha v+u)^{2^i}+v^{2^i}\right)y=0, \label{equ:hmdf2}
  \end{numcases}
  has at most 4 solutions over $\GF{k}^2$.\par

  \noindent(I) In the case of $\alpha^{2^i}(\alpha u+v)+u=0$. Then $\alpha u+v\neq0$, otherwise, we have $u=v=0$, which is impossible. And also $\alpha(\alpha u+v)^{2^i}+u^{2^i}=\alpha(\alpha u+v)^{2^i}+\left(\alpha^{2^i}(\alpha u+v)\right)^{2^i}=(\alpha^{2^{2i}}+\alpha)(\alpha u+v)^{2^i}\neq 0$ since $\alpha\neq 0,1$, $k$ odd and $\gcd(i,k)=1$. Then \eqref{equ:hmdf1} can be written
  \begin{equation}\label{equ:hmdf3}
    x=\frac{\alpha u+v}{(\alpha^{2^{2i}}+\alpha)(\alpha u+v)^{2^i}}y^{2^i}+\frac{1}{\alpha^{2^{2i}}+\alpha}y.
  \end{equation}\par

  If $\alpha v+u=0$, \eqref{equ:hmdf2} becomes $vy^{2^i}+v^{2^i}y=0$. Note that we must have $v\neq 0$, therefore, $y=0$ or $y=v$. \eqref{equ:hmdf3} in $x$ have only one solution with respect to each of $y$. Hence, the total number of solutions is equal to 2.\par

  If $\alpha v+u\neq 0$. We replace \eqref{equ:hmdf2} by \eqref{equ:hmdf3} and get
  \[
   A_1 y^{2^{2i}}+A_2 y^{2^i}+A_3 y=0,
  \]
  where
  \[
    \begin{split}
      A_1 & =  \frac{(\alpha v+u)(\alpha u+v)^{2^i}}{(\alpha^{2^{2i}}+\alpha)^{2^i}(\alpha u+v)^{2^{2i}}}, \\
      A_2 & =  \frac{(\alpha v+u)(\alpha u+v)^{2^i}}{(\alpha^{2^{2i}}+\alpha)^{2^i}(\alpha u+v)^{2^{2i}}}+\frac{(\alpha v+u)^{2^i}(\alpha u+v)}{(\alpha^{2^{2i}}+\alpha)^{2^i}(\alpha u+v)^{2^{2i}}}+\alpha^{2^i}(\alpha v+u)+v,\\
      A_3 & =  \frac{(\alpha v+u)^{2^i}(\alpha u+v)^{2^i}}{(\alpha^{2^{2i}}+\alpha)(\alpha u+v)^{2^i}}+\alpha(\alpha v+u)^{2^i}+v^{2^i}.
    \end{split}
  \]
  Notice that $A_1\neq 0$, hence, according to Corollary \ref{cor:4sol2}, the above equation in $y$ has at most 4 solutions. From \eqref{equ:hmdf3}, the solution $x$ is uniquely determined by $y$. Hence, the total number of solutions is at most equal to 4.\par

  \noindent(II) In the case of $\alpha^{2^i}(\alpha u+v)+u\neq 0$.\par

  If $\alpha v+u=0$, \eqref{equ:hmdf2} becomes $vy^{2^i}+v^{2^i}y=0$. Note that we must have $v\neq 0$, therefore, $y=0$ or $y=v$. For each of the solutions $y$, \eqref{equ:hmdf1} in $x$ has at most 2 solutions. Hence, the total number of solutions is at most equal to 4.\par

  If $\alpha v+u\neq 0$. We add \eqref{equ:hmdf1} multiplied by $\alpha v+u$ to \eqref{equ:hmdf2} multiplied by $\alpha^{2^i}(\alpha u+v)+u$, from which we eliminate $x^{2^i}$ and get
  \begin{equation}{\label{equ:hmdf4}}
    A_4 x+A_5 y^{2^i}+A_6 y=0,
  \end{equation}
  where
  \[
    \begin{split}
      A_4 & =  (\alpha v+u)\left(\alpha(\alpha u+v)^{2^i}+u^{2^i}\right)+(\alpha v+u)^{2^i}\left(\alpha^{2^i}(\alpha u+v)+u\right), \\
      A_5 & =  (\alpha v+u)(\alpha u+v)+\left(\alpha^{2^i}(\alpha u+v)+u\right)\left(\alpha^{2^i}(\alpha v+u)+v\right),\\
      A_6 & =  (\alpha v+u)(\alpha u+v)^{2^i}+\left(\alpha^{2^i}(\alpha u+v)+u\right)\left(\alpha(\alpha v+u)^{2^i}+v^{2^i}\right).
    \end{split}
  \]
  According to Lemma \ref{lem:dfudp}, not all of $A_4,A_5,A_6$ are equal to 0.\par

  If $A_4=0$, from \eqref{equ:hmdf4}, $y$ has at most 2 solutions. For each of the solutions $y$, \eqref{equ:hmdf1} in $x$ has at most 2 solutions. Hence, the total number of solutions is at most equal to 4.\par

  If $A_4\neq 0$ and $A_5=A_6=0$, then $x=0$. Since not both of $\alpha^{2^i}(\alpha v+u)+v$ and $\alpha(\alpha v+u)^{2^i}+v^{2^i}$ equal to 0, otherwise, we get $(\alpha^{2^{2i}}+\alpha)(\alpha v+u)^{2^i}=0$, which is impossible. Replace \eqref{equ:hmdf2} by $x=0$, we obtain a equation in $y$ with coefficient not all zero with at most 2 solutions. Hence, the total number of solutions is at most equal to 2.\par

  If $A_4\neq 0,A_5=0,A_6\neq 0$, we replace \eqref{equ:hmdf1} by \eqref{equ:hmdf4} and get
  \begin{equation}\label{equ:hmdf5}
    A_7y^{2^i}+A_8y=0,
  \end{equation}
  where
  \[
    \begin{split}
    A_7 & = \left(\alpha^{2^i}(\alpha u+v)+u\right)\left(\frac{A_6}{A_4}\right)^{2^i}+(\alpha u+v),\\
    A_8 & = \left(\alpha(\alpha u+v)^{2^i}+u^{2^i}\right)\frac{A_6}{A_4}+(\alpha u+v)^{2^i}.
    \end{split}
  \]
  We claim that $A_8\neq 0$, otherwise,
  \[
    \left(\alpha(\alpha u+v)^{2^i}+u^{2^i}\right)A_6=(\alpha u+v)^{2^i}A_4.
  \]
  Replace above equation by the expressions $A_4$ and $A_6$, recall that $\alpha^{2^i}(\alpha u+v)+u\neq 0$, and we get
  \begin{equation}\label{equ:hmdf6}
      (\alpha v+u)(\alpha u+v)=(\beta(\alpha u+v)+u)(\beta(\alpha v+u)+v),
  \end{equation}
  where $\alpha=\beta^{2^i}$. If $u=0$, then $v\neq 0$. From $A_5=0$ and \eqref{equ:hmdf6}, we obtain $\beta^{2^i+2}+\beta^{2^i}+\beta=0$ and $\alpha^{2^{i+1}+1}+\alpha^{2^i}+\alpha=0$. We add the first equation raised up to $2^i$ to the second equation and get $\alpha^{2^i+1}(\alpha^{2^i}+\alpha)=0$, which is impossible. The case $v=0,u\neq 0$ is identical. Hence, we can assume the $u\neq 0,v\neq 0$. Observe that equations $A_5=0$ and \eqref{equ:hmdf6} are the same with \eqref{equ:auv02} and \eqref{equ:auv04}, then according the proof of Lemma \ref{lem:dfudp}, this is impossible, which means $A_8\neq 0$. Therefore, \eqref{equ:hmdf5} in $y$ has at most 2 solutions. For each of the solutions, \eqref{equ:hmdf4} in $x$ has only one solution. Hence, \eqref{equ:odfuv} has at most 4 solutions.\par

  If $A_4\neq 0$ and $A_5\neq 0$, we replace \eqref{equ:hmdf2} by \eqref{equ:hmdf4} and get
  \begin{equation}{\label{equ:hmdf7}}
    A_9 y^{2^{2i}}+A_{10} y^{2^i}+A_{11} y=0,
  \end{equation}
  where $A_9=(\alpha v+u)(\frac{A_5}{A_4})^{2^i}\neq 0$ and $A_{10}, A_{11}$ are expressions of $\alpha,u,v$. According to Corollary \ref{cor:4sol2}, \eqref{equ:hmdf7} in $y$ has at most 4 solutions. For each solution, \eqref{equ:hmdf4} in $x$ has only one solution. Hence, \eqref{equ:odfuv} has at most 4 solutions.\par

  Therefore, \eqref{equ:odfuv} has at most 4 solutions, meaning that the differential uniformity of $\Vb_e^{\alpha}$ is at most 4. We complete the proof. \QED
\end{prf}

\begin{rmk}
  In \cite{Butterfly}, it is proved that the 6-bit APN permutation described by Dillon et al. is affine equivalent to the butterfly structures $\Hb_6^2$. However, when $k>3$, the pair $(e,\alpha)$ such that $\Hb_e^{\alpha}$ is APN has not been found. The authors verified experimentally that butterfly structures are never differentially 4-uniform for $k=4,8,10$, while cases $k=6$ there does exist.
\end{rmk}

\subsection{Algebraic Degree}\label{sbs:aldeg}

When $e=2^i+1$, the left and right side of $\Vb_e^{\alpha}(x,y)$ are equal to $(\alpha x+y)^{2^i+1}+x^{2^i+1}$, $(x+\alpha y)^{2^i+1}+y^{2^i+1}$ respectively. It is obvious that it is quadratic. Now we consider the open butterfly structure $\Hb_e^{\alpha}$. The following result is needed.\par

\begin{lem}[\cite{DefDiff}]\label{lem:degre}
  Suppose $k$ is odd integer and $\gcd(i,k)=1$. Then the compositional inverse of $x^{2^i+1}$ over $\GF{k}$ is $x^t$, where $t=\sum_{j=0}^{\frac{k-1}{2}}2^{2ji} \mod (2^k-1)$. Its algebraic degree is $\frac{k+1}{2}$.
\end{lem}

The right side of the output of the open butterfly $\Hb_e^{\alpha}$ is equal to $(x+y^{2^i+1})^{\frac{1}{2^i+1}}+\alpha y$. According to Lemma \ref{lem:degre}, we deduce from this expression that $(x+y^{2^i+1})^{\frac{1}{2^i+1}}$ is equal to $\prod_{j=0}^{\frac{k-1}{2}}(x+y^{2^i+1})^{2^{2ji}}$. This sum can be developed as follows:
\[
(x+y^{2^i+1})^{\frac{1}{2^i+1}}=\sum_{J\subseteq[0,(k-1)/2]}\underbrace{\prod_{j\in J}y^{(2^i+1)\cdot 2^{2ji}}}_{\deg \leq 2|J|}\underbrace{\prod_{j\in\overline{J}}x^{2^{2ji}}}_{\deg =(k+1)/2-|J|},
\]
where $\overline{J}$ is the complement of $J$ in $[0,(k-1)/2]$. The algebraic degree of each term in this sum is at most equal to $(k+1)/2+|J|$. If $\overline{J}=\emptyset$ then $x$ is absent from the term so that the algebraic degree of this term is equal to $w_2\left(\sum_{j=0}^{(k-1)/2}(2^i+1)\cdot 2^{2ji}\right)=w_2\left(1\right)=1$. If $|\overline{J}|=1$, then the algebraic degree of this term is at most $(k-1)/2+(k+1)/2=k$. Moveover, if $\overline{J}=\{0\}$, then the term is equal to $x(y^{2^i})^{-1}$ which has algebraic degree $1+(k-1)=k$. If $|\overline{J}|>1$, the degree of these terms is at most equal to $(k-3)/2+(k+1)/2=k-1$. Thus, the right side of the output has an algebraic degree equal to $k$.\par

The left side is equal to
\[
\left(y+\alpha\left(\left(x+y^{2^i+1}\right)^{\frac{1}{2^i+1}}+\alpha y\right)\right)^{2^i+1}+\left(\left(x+y^{2^i+1}\right)^{\frac{1}{2^i+1}}+\alpha y\right)^{2^i+1}.
\]
The terms of highest algebraic degree in this equation are of the shape $y^{2^i}(x+y^{2^i+1})^{\frac{1}{2^i+1}}$ and $y(x+y^{2^i+1})^{2^i\cdot{\frac{1}{2^i+1}}}$. We still have
\[
y^{2^i}(x+y^{2^i+1})^{\frac{1}{2^i+1}}=\sum_{J\subseteq[0,(k-1)/2]}\underbrace{y^{2^i}\cdot\prod_{j\in J}y^{(2^i+1)\cdot 2^{2ji}}}_{\deg \leq 2|J|+1}\underbrace{\prod_{j\in\overline{J}}x^{2^{2ji}}}_{\deg =(k+1)/2-|J|},
\]
so that the algebraic degree of each term is at most equal to $(k+1)/2+|J|+1$. If $|J|=(k+1)/2$, i.e.,$\overline{J}=\emptyset$, then $x$ is absent from the term so that the algebraic degree of this term is equal to $w_2\left(2^i+\sum_{j=0}^{(k-1)/2}(2^i+1)\cdot 2^{2ji}\right)=w_2(2^i+1)=2$. If $|J|\leq (k-1)/2$, then the algebraic degree of these terms is at most equal to $(k+1)/2+|J|+1\leq k+1$. For $J=[1,(k-1)/2]$, then the algebraic degree of this term is $1+w_2\left(2^i+(2^i+1)\sum_{j=1}^{(k-1)/2}2^{2ji}\right)=1+w_2\left(\sum_{j=0}^{(k-1)/2}2^{ji}\right)=k+1$. The terms $y(x+y^{2^i+1})^{2^i\cdot{\frac{1}{2^i+1}}}$ are treated similarly. Hence the left side of the output has algebraic degree $k+1$.\par

\begin{thm}
  Let $n$ is an odd integer and $\gcd(i,k)=1$. For any $0\leq t\leq k-1$, $\alpha\in\GF{k}$, and $\alpha\neq 0,1$, let $\Hb_{e}^{\alpha}$ and $\Vb_{e}^{\alpha}$ be the open and closed $2k$-bit butterflies structure with exponent $e=(2^i+1)\times 2^t$ and coefficient $\alpha$. Then
  \begin{enumerate}[(1)]
    \item The algebraic degree of $\Vb_e^{\alpha}$ is 2.
    \item The left and right side of the output of $\Hb_e^{\alpha}$ have algebraic degree $k+1$ and $k$ respectively.
  \end{enumerate}
\end{thm}

\subsection{Nonlinearity}\label{sbs:nonli}

In this section, we consider the nonlinearity, the following lemma is needed.\par

\begin{lem}\label{lem:nldep}
  Let $n$ is an odd integer and $\gcd(i,k)=1$. Then for any $\alpha\in\GF{k}$ with $\alpha\neq 0,1 $, the following system of equations in variables $c,d$
  \begin{numcases}{}
     \left(\alpha^{2^i+1}c+c+d\right)\left(\alpha^{2^i}c+\alpha d\right)^{2^i}+\left(\alpha^{2^i+1}c+c+d\right)^{2^i}\left(\alpha c+\alpha^{2^i}d\right)=0,\label{equ:acd01}\\
     \left(\alpha c+\alpha^{2^i}d\right)^{2^i}\left(\alpha^{2^i}c+\alpha d\right)^{2^i}+\left(\alpha^{2^i+1}c+c+d\right)^{2^i}\left(\alpha^{2^i+1}d+c+d\right)^{2^i}=0, \label{equ:acd02}\\
     \left(\alpha^{2^i}c+\alpha d\right)\left(\alpha^{2^i}c+\alpha d\right)^{2^i}+\left(\alpha^{2^i+1}c+c+d\right)^{2^i}\left(\alpha^{2^i+1}d+c+d\right)=0 \label{equ:acd03}
\end{numcases}
  holds over $\GF{k}$ if and only if $c,d$ satisfying $\alpha^{2^i+1}c+c+d=0$ and $\alpha^{2^i}c+\alpha d=0$.
\end{lem}

\begin{prf}
  The sufficiency is obvious. Now we show the necessary.\par

  If $\alpha^{2^i+1}c+c+d=0,\alpha^{2^i}c+\alpha d\neq 0$, which contradicts \eqref{equ:acd03}. If $\alpha^{2^i+1}c+c+d\neq 0,\alpha^{2^i}c+\alpha d=0$, from \eqref{equ:acd01} we must have $\alpha c+\alpha^{2^i}d=0$, which implies that $(\alpha^{2^i}+ \alpha)(c+d)=0$. Hence, $c=d=0$, which contradicts $\alpha^{2^i+1}c+c+d\neq 0$.\par

  Now we always assume that $\alpha^{2^i+1}c+c+d\neq 0,\alpha^{2^i}c+\alpha d\neq 0$. Then $\alpha c+\alpha^{2^i}d\neq 0$ and $\alpha^{2^i+1}d+c+d\neq 0$, otherwise, we must have $\alpha^{2^i+1}c+c+d= 0$ or $\alpha^{2^i}c+\alpha d= 0$ from \eqref{equ:acd01} and \eqref{equ:acd03}.\par

  If $c=0$, and replace it into \eqref{equ:acd02}, we obtain $d^{2^{i+1}}=0$, which contradicts the hypothesis. The case $d=0$ is identical. Therefore, we also assume that $c\neq 0,d\neq 0$. \par

  According to \eqref{equ:acd01} and \eqref{equ:acd02}, we have
  \begin{equation*}
    \frac{\alpha^{2^i+1}c+c+d}{\alpha c+\alpha^{2^i}d}=\frac{\left(\alpha^{2^i+1}c+c+d\right)^{2^i}}{\left(\alpha^{2^i}c+\alpha d\right)^{2^i}}=\frac{\left(\alpha c+\alpha^{2^i}d\right)^{2^i}}{\left(\alpha^{2^i+1}d+c+d\right)^{2^i}},
  \end{equation*}
  and obtain the following equation
  \[
  \left(\alpha^{2^i+1}c+c+d\right)\left(\alpha^{2^i+1}d+c+d\right)^{2^i}=\left(\alpha c+\alpha^{2^i}d\right)\left(\alpha c+\alpha^{2^i}d\right)^{2^i}.
  \]
  To simplify expressions, we denote $c=\lambda d,\lambda\neq 0$, and use the notation $\beta=\alpha^{2^i},\beta\neq 0,1$. The above equation can be rewritten as
  \begin{equation}\label{equ:acd04}
    \lambda^{2^i+1}+(\beta^2+1)\lambda^{2^i}+(\alpha\beta^{2^i+2}+\beta^{2^i+1}+\alpha\beta^{2^i}+\alpha\beta+1)\lambda+1=0.
  \end{equation}\par

  Similarly, \eqref{equ:acd03} can be rewritten as
  \begin{equation}\label{equ:acd05}
    \lambda^{2^i+1}+(\alpha\beta^{2^i+2}+\beta^{2^i+1}+\alpha\beta^{2^i}+\alpha\beta+1)\lambda^{2^i}+(\beta^2+1)\lambda+1=0.
  \end{equation}
  We add \eqref{equ:acd04} to \eqref{equ:acd05} and get
  \begin{equation}\label{equ:acd06}
    (\alpha\beta^{2^i+2}+\beta^{2^i+1}+\alpha\beta^{2^i}+\alpha\beta+\beta^2)\lambda^{2^i}+ (\alpha\beta^{2^i+2}+\beta^{2^i+1}+\alpha\beta^{2^i}+\alpha\beta+\beta^2)\lambda=0.
  \end{equation}
  Recall that $\alpha\beta^{2^i+2}+\beta^{2^i+1}+\alpha\beta^{2^i}+\alpha\beta+\beta^2=\alpha^{2^{2i}+2^{i+1}+1}+\alpha^{2^{2i}+2^i}+ \alpha^{2^{2i}+1}+\alpha^{2^{i+1}}+\alpha^{2^i+1}\neq 0$, then \eqref{equ:acd06} becomes $\lambda^{2^i}+\lambda=0$. Hence, $\lambda =1$, which means $c=d$. Replace \eqref{equ:acd02} by $c=d$ and we have $\left((\alpha +1)^{2^i+1}+1\right)^2 c^2=0$, which is a contradiction since $\alpha\neq 0$ and $c\neq 0$.\par

  We complete the proof.\QED

\end{prf}

\begin{lem}\label{lem:4sols}
  Let $n$ is an odd integer and $\gcd(i,k)=1$. Then for any $(c,d)\in\GF{k}\times\GF{k}$ with $(c,d)\neq (0,0)$, the following system of equations in variables $u,v$
  \begin{numcases}{\hspace{-7mm}}
    \left(\alpha^{2^i+1}c+c+d\right)^{2^i}u^{2^{2i}}+\left(\alpha^{2^i+1}c+c+d\right)u+ \left(\alpha c+\alpha^{2^i} d\right)^{2^i}v^{2^{2i}}+\left(\alpha^{2^i}c+\alpha d\right)v=0, \label{equ:4sol1}\\
    \left(\alpha^{2^i}c+\alpha d\right)^{2^i}u^{2^{2i}}+\left(\alpha c+\alpha^{2^i} d\right)u+ \left(\alpha^{2^i+1}d+c+d\right)^{2^i}v^{2^{2i}}+\left(\alpha^{2^i+1}d+c+d\right)v=0 \label{equ:4sol2}
  \end{numcases}
  has at most 4 solutions over $\GF{k}^2$.
\end{lem}

\begin{prf}
  \noindent (I) In the case of $\alpha^{2^i+1}c+c+d=0$, \eqref{equ:4sol1} becomes
  \begin{equation}\label{equ:4sol3}
    \left(\alpha c+\alpha^{2^i} d\right)^{2^i}v^{2^{2i}}+\left(\alpha^{2^i}c+\alpha c\right)v=0.
  \end{equation}\par

  If $\alpha c+\alpha^{2^i}d=0$, then $\alpha^{2^i}c+\alpha d\neq 0$, otherwise, we can obtain $c=d=0$. Hence, $v=0$, replace it into \eqref{equ:4sol2} and we get $u=0$. So we have only solution $(0,0)$.\par

  If $\alpha^{2^i}c+\alpha d=0$, then $\alpha c+\alpha^{2^i}d\neq 0$. Similarly, we have only solution $(0,0)$.\par

  If $\alpha^{2^i}c+\alpha d\neq 0, \alpha c+\alpha^{2^i}d\neq 0$, then \eqref{equ:4sol3} in $v$ has 2 solutions. For each one $v$ of the solutions, \eqref{equ:4sol2} in $x$ has at most 2 solutions. Hence, the total numbers of solutions is at most equal to 4.\par

  \noindent (II) In the case of $\alpha^{2^i+1}c+c+d\neq 0$.\par

  If $\alpha^{2^i}c+\alpha d=0$, then $\alpha c+\alpha^{2^i}d\neq 0$. \eqref{equ:4sol2} becomes
  \begin{equation}\label{equ:4sol4}
    \left(\alpha c+\alpha^{2^i} d\right)u+ \left(\alpha^{2^i+1}d+c+d\right)^{2^i}v^{2^{2i}}+\left(\alpha^{2^i+1}d+c+d\right)v=0.
  \end{equation}
  When $\alpha^{2^i+1}d+c+d=0$, then $u=0$. Replace it into \eqref{equ:4sol1} and we have only one solution $(0,0)$. When $\alpha^{2^i+1}d+c+d\neq 0$, Replace \eqref{equ:4sol1} by \eqref{equ:4sol4} and we obtain
  \[
  B_1v^{2^{4i}}+B_2v^{2^{2i}}+B_3v=0,
  \]
  where $B_1=\frac{\left(\alpha^{2^i+1}c+c+d\right)\left(\alpha^{2^i+1}d+c+d\right)^{2^{3i}}}{\left(\alpha c+\alpha^{2^i} d\right)^{2^{2i}}}\neq 0$, $B_2,B_3$ are expressions of $\alpha,c,d$. According to Corollary \ref{cor:4sol4}, the above equation in $v$ has at most 4 solutions. For each solution, \eqref{equ:4sol4} in $u$ has only one solution. Hence, the total numbers of solutions is at most equal to 4.\par

  If $\alpha^{2^i}c+\alpha d\neq 0$, we add \eqref{equ:4sol1} multiplied by $\left(\alpha^{2^i}c+\alpha d\right)^{2^i}$ to \eqref{equ:4sol2} multiplied by $\left(\alpha^{2^i+1}c+c+d\right)^{2^i}$ to eliminate $u^{2^{2i}}$ and get
  \begin{equation}\label{equ:4sol5}
    B_4u+B_5v^{2^{2i}}+B_6v=0,
  \end{equation}
  where
  \[
  \begin{split}
    B_4 & = \left(\alpha^{2^i+1}c+c+d\right)\left(\alpha^{2^i}c+\alpha d\right)^{2^i}+\left(\alpha^{2^i+1}c+c+d\right)^{2^i}\left(\alpha c+\alpha^{2^i}d\right),\\
    B_5 & = \left(\alpha c+\alpha^{2^i}d\right)^{2^i}\left(\alpha^{2^i}c+\alpha d\right)^{2^i}+\left(\alpha^{2^i+1}c+c+d\right)^{2^i}\left(\alpha^{2^i+1}d+c+d\right)^{2^i},\\
    B_6 & = \left(\alpha^{2^i}c+\alpha d\right)\left(\alpha^{2^i}c+\alpha d\right)^{2^i}+\left(\alpha^{2^i+1}c+c+d\right)^{2^i}\left(\alpha^{2^i+1}d+c+d\right).
  \end{split}
  \]
  According to Lemma \ref{lem:nldep}, not all of $B_4,B_5,B_6$ are equal to 0.\par

  If $B_4=0$, from \eqref{equ:4sol5}, $v$ has at most 2 solutions. For each of the solutions $v$, \eqref{equ:4sol1} in $u$ has at most 2 solutions. Hence, the total number of solutions is at most equal to 4.\par

  If $B_4\neq 0$ and $B_5=B_6=0$, then $u=0$. Recall that not both of $\alpha^{2^i}c+\alpha c$ and $\alpha c+\alpha^{2^i}d$ are equal to 0, Replace \eqref{equ:4sol1} by $u=0$, we obtain a equation in $v$ with coefficient not all zero with at most 2 solutions. Hence, the total number of solutions is at most equal to 2.\par

  If $B_4\neq 0$ and $B_5=0,B_6\neq 0$, then replace \eqref{equ:4sol5} into \eqref{equ:4sol1} and obtain
  \[
  B_7v^{2^{2i}}=0
  \]
  where $B_7= \left(\alpha^{2^i}c+\alpha d\right)^{2^i}B_6^{2^{2i}}+\left(\alpha^{2^i+1}d+c+d\right)^{2^i}B_4^{2^{2i}}$. With a tedious verification (See Appendix) we have $B_7\neq 0$. Hence, $v=0$, which implies that $u=0$. We have only one solution.\par

  If $B_4\neq 0$ and $B_5\neq 0$, we replace \eqref{equ:4sol2} by \eqref{equ:4sol5} and get
  \begin{equation}{\label{equ:4sol6}}
    B_8 v^{2^{4i}}+B_9 v^{2^{2i}}+B_{10} v=0,
  \end{equation}
  where $B_8=\left(\alpha^{2^i}c+\alpha d\right)^{2^i}\left(\frac{B_5}{B_4}\right)^{2^{2i}}\neq 0$ and $B_9, B_{10}$ are expressions of $\alpha,c,d$. According to Corollary \ref{cor:4sol4}, \eqref{equ:4sol6} in $v$ has at most 4 solutions. For each solution, \eqref{equ:4sol5} in $u$ has only one solution. Hence, the total number of solutions is at most equal to 4.\par

  We complete the proof.\QED
\end{prf}

\begin{thm}\label{thm:nolin}
  Let $n$ is an odd integer and $\gcd(i,k)=1$. For any $0\leq t\leq k-1$, $\alpha\in\GF{k}$, and $\alpha\neq 0,1$, let $\Hb_{e}^{\alpha}$ and $\Vb_{e}^{\alpha}$ be the open and closed $2k$-bit butterflies structure with exponent $e=(2^i+1)\times 2^t$ and coefficient $\alpha$. Then the nonlinearity of both $\Hb_{e}^{\alpha}$ and $\Vb_{e}^{\alpha}$ is $2^{2k-1}-2^k$. Furthermore, their Walsh spectrum are $\{0,\pm 2^k,\pm 2^{k+1}\}$.
\end{thm}

\begin{prf}
  As $\Hb_{e}^{\alpha}$ and $\Vb_{e}^{\alpha}$ are CCZ-equivalent, they have the same nonlinearity and walsh spectrum. It is sufficient to prove that the nonlinearity of $\Vb_{e}^{\alpha}$ is $2^{2k-1}-2^k$. Besides, the functions $\Vb_{e}^{\alpha}$ with exponent $e=(2^i+1)\times 2^t$ is affine equivalent to functions $\Vb_{e}^{\alpha}$ with exponent $e=2^i+1$. Thus it is sufficient to consider the case where the exponent is equal to $e=2^i+1$.\par

  Let $a,b,c,d\in\GF{k}$, and $(c,d)\neq(0,0)$. Then we have
  \[
  \begin{split}
  \mathcal{W}_F((a,b),(c,d)) & = \sum_{x,y\in\GF{k}}(-1)^{\Tr(c(\alpha x+y)^{2^i+1}+cx^{2^i+1}+d(x+\alpha y)^{2^i+1}+dy^{2^i+1}+ax+by)}\\
                             & = \sum_{x,y\in\GF{k}}(-1)^{f(x,y)},
  \end{split}
  \]
  where
  \[
  \begin{split}
  f(x,y) = & \Tr\left((\alpha^{2^i+1}c+c+d)x^{2^i+1}+(\alpha^{2^i+1}d+c+d)y^{2^i+1}\right.\\
           &  \qquad \left.+(\alpha^{2^i}c+\alpha d)x^{2^i}y+(\alpha c+\alpha^{2^i}d)xy^{2^i}+ax+by\right).
  \end{split}
  \]\par

  Using the fact that $\Tr(x)=\Tr(x^{2^i})$, we deduce the following representation
  \[
  \begin{split}
     & f(x,y)+f(x+u,y+v)\\
  =  & \Tr\left[\left((\alpha^{2^i+1}c+c+d)^{2^i}u^{2^{2i}}+(\alpha^{2^i+1}c+c+d)u+(\alpha c+\alpha^{2^i} d)^{2^i}v^{2^{2i}}+ (\alpha^{2^i}c+\alpha d)v\right)x^{2^i} \right.\\
     & \hspace{5mm}\left((\alpha^{2^i}c+\alpha d)^{2^i}u^{2^{2i}}+(\alpha c+\alpha^{2^i}d)u+ \left.(\alpha^{2^i+1}d+c+d)^{2^i}v^{2^{2i}}+(\alpha^{2^i+1}d+c+d)v\right)y^{2^i}\right]\\
     & + f(u,v),
  \end{split}
  \]
  then it holds that
  \[
  \begin{split}
     \mathcal{W}_F^2((a,b),(c,d)) & = \sum_{x,y\in\GF{k}}(-1)^{f(x,y)}\times\sum_{u,v\in\GF{k}}(-1)^{f(x+u,y+v)} \\
                                  & = \sum_{x,y,u,v\in\GF{k}}(-1)^{f(x,y)+f(x+u,y+v)}\\
                                  & = 2^{2k}\sum_{u,v\in S(c,d)}(-1)^{f(u,v)},
  \end{split}
  \]
  where $R(c,d)$ is the solution set of the following system of equations with variables $u,v$
  \[
  \begin{cases}
    \left(\alpha^{2^i+1}c+c+d\right)^{2^i}u^{2^{2i}}+\left(\alpha^{2^i+1}c+c+d\right)u+ \left(\alpha c+\alpha^{2^i} d\right)^{2^i}v^{2^{2i}}+\left(\alpha^{2^i}c+\alpha d\right)v=0, \\
    \left(\alpha^{2^i}c+\alpha d\right)^{2^i}u^{2^{2i}}+\left(\alpha c+\alpha^{2^i} d\right)u+ \left(\alpha^{2^i+1}d+c+d\right)^{2^i}v^{2^{2i}}+\left(\alpha^{2^i+1}d+c+d\right)v=0.
  \end{cases}
  \]
  Denote $m=\dim_{F_2}R(c,d)$, according Lemma \ref{lem:4sols}, $0\leq m\leq 2$. Notice that $f(x,y)+f(x+u,y+v)=f(u,v)$ for $(u,v)\in R(c,d)$ and $(x,y)\in\GF{k}^2$, which means $f(u,v)$ is linear over $R(c,d)$. Since $(0,0)\in R(c,d)$, therefore, $f(u,v)$ is a balanced or constant 0 over $R(c,d)$. Thus
  \[
  \mathcal{W}_F^2((a,b),(c,d))=
  \begin{cases}
    2^{2k+m} & f(u,v)=0 \text{ over } R(c,d),\\
    0        & \text{otherwise}.
  \end{cases}
  \]
  As $\mathcal{W}_F((a,b),(c,d))$ is an integer, $m$ must be even, i.e., $m=0$ or $m=2$. Hence, $\mathcal{W}_F((a,b),(c,d))\in\{0,\pm 2^k,\pm 2^{k+1}\}$.\par

  Since $\Hb_e^{\alpha}$ is a permutation over $\GF{k}^2$, $\mathcal{W}_F((0,0),(c,d))=0$ for any $(c,d)\in\GF{k}^2$ with $(c,d)\neq (0,0)$, which means $0\in\Lambda_F$. Besides, we also have $\pm 2^{k+1}\in\Lambda_F$, otherwise, according to Parseval’s equality we must have $\mathcal{W}_F((a,b),(c,d))=\pm 2^k$ for any $(a,b),(c,d)\in\GF{k}^2$ with $(c,d)\neq (0,0)$, which is impossible. If $\pm 2^k\notin\Lambda_F$, according to Lemma \ref{lem:aldeg}, the algebraic degree is at most equal to $2k-(k+1)+1=k$, which contradicts the algebraic degree of $\Hb_e^{\alpha}$ is $k+1$. Therefore, $\Lambda_F=\{0,\pm 2^{k},\pm 2^{k+1}\}$, and the nonlinearity $\mathcal{NL}(F)=2^{2k-1}-2^k$.\par

  We complete the proof.\QED

\end{prf}

\begin{rmk}
  Recall that the Walsh spectrum of Gold functions are $\{0, \pm2^{k+1}\}$, which is different from that of butterfly structures. Hence, the butterfly structures $\Hb_e^{\alpha}$ and $\Vb_e^{\alpha}$ is CCZ-inequivalent to the Gold functions. Besides, in the proof of Lemma \ref{lem:4sols}, there exists some cases that the solution sets $R(c,d)$ has only one solution $(0,0)$ (e.g. the case of $\alpha c+\alpha^{2^i}d=0$ and $\alpha^{2^i}c+\alpha c\neq 0$), namely, $m=0$. Hence, we also have $\pm 2^k\in\Lambda_F$. From the proof of above theorem, we have actually $m=0$ or $m=2$, meaning that the system of equations in Lemma \ref{lem:4sols} has one solution or 4 solutions.
\end{rmk}


\section{Butterfly Structure with $\alpha=1$}\label{sec:trivl}

In this section, we study the butterflies with trivial coefficient $\alpha=1$. We show that $\Vb_e^1$ is also a permutation in section \ref{sbs:bijec}. In section \ref{sbs:crypp} we consider other cryptographic properties.

\subsection{The bijective of closed butterfly structure}\label{sbs:bijec}

When $\alpha =1, e=2^i+1$, the closed butterfly $\Vb_e^1$ becomes
\[
\Vb_e^1(x,y)=((x+y)^{2^i+1}+x^{2^i+1},(x+y)^{2^i+1}+y^{2^i+1}).
\]
Then we have the following result.

\begin{thm}
  Let $n$ is an odd integer and $\gcd(i,k)=1$. For any $0\leq t\leq k-1$, let $\Vb_{e}^1$ be the closed $2k$-bit butterflies structure with exponent $e=(2^i+1)\times 2^t$. Then $\Vb_{e}^1(x,y)$ is a permutation over $\GF{k}^2$.
\end{thm}

\begin{prf}
  Similarly, we consider the case $e=2^i+1$. For any $u,v\in\GF{k}$ and $(u,v)\neq (0,0)$. It is sufficient to show that
  \[
  \Vb_e^1(x,y)+\Vb_e^1(x+u,y+v)=(0,0),
  \]
  namely, the system of equations
  \[
  \begin{cases}
    vx^{2^i}+v^{2^i}x+(u+v)y^{2^i}+(u+v)^{2^i}y=(u+v)^{2^i+1}+u^{2^i+1}, \\
    (u+v)x^{2^i}+(u+v)^{2^i}x+uy^{2^i}+u^{2^i}y=(u+v)^{2^i+1}+v^{2^i+1}.
  \end{cases}
  \]
  has no solution over $\GF{k}^2$. We consider the following equivalent system of equations
  \begin{numcases}{}
    ux^{2^i}+u^{2^i}x+vy^{2^i}+v^{2^i}y=u^{2^i+1}+v^{2^i+1}, \label{equ:perm1}\\
    (u+v)x^{2^i}+(u+v)^{2^i}x+uy^{2^i}+u^{2^i}y=(u+v)^{2^i+1}+v^{2^i+1}. \label{equ:perm2}
  \end{numcases}\par

  First, if $u=0$, then $v\neq 0$. So equation \eqref{equ:perm1} becomes
  \[
  vy^{2^i}+v^{2^i}y=v^{2^i+1},
  \]
  which, in fact, is equivalent to $(v+y)^{2^i+1}=y^{2^i+1}$. Therefore, equation (\ref{equ:perm1}) has no solution over $\GF{k}^2$ since $x^{2^i+1}$ is a permutation over $\GF{k}$.\par

  The case of $u\neq 0, v=0$ and $u=v\neq 0$ can be proved similarly.\par

  Next, we suppose that $u\neq 0, v\neq 0$, and $u\neq v$. To eliminate $y^{2^i}$, we add equation (\ref{equ:perm1}) multiplied by $u$ to equation (\ref{equ:perm2}) multiplied by $v$ and get
  \[
  y=\frac{1}{C_2}(C_1x^{2^i}+C_3x+C_1u^{2^i}),
  \]
  where
  \[
  \begin{split}
    C_1 & =u^2+uv+v^2,\\
    C_2 & =u^{2^i}v+uv^{2^i},\\
    C_3 & =u^{2^i+1}+u^{2^i}v+v^{2^i+1}.
  \end{split}
  \]
  It is easy to see that $C_1\neq 0$ and $C_2\neq 0$ since that $k$ is odd, $\gcd(i,k)=1, u\neq 0, v\neq 0$, and $u\neq v$. Substitute the above equation to equation \eqref{equ:perm1} and multiply both sides by $(C_2)^{2^i+1}$, then we obtain
  \begin{equation*}
    \begin{split}
      & vC_2C_1^{2^i}x^{2^{2i}}+\left(uC_2^{2^i+1}+(vC_2)^{2^i}C_1+vC_2C_3^{2^i}\right)x^{2^i}+ \left(u^{2^i}C_2^{2^i+1}+(vC_2)^{2^i}C_3\right)x \\
       =\ & vC_2C_1^{2^i}u^{2^{2i}}+(vC_2)^{2^i}C_1u^{2^i}+C_2^{2^i+1}(u^{2^i+1}+v^{2^i+1}).
    \end{split}
  \end{equation*}\par

  We simplify respectively the coefficient of each term of the above equation, and finally get
  \[
   C_2x^{2^{2i}}+(u^{2^{2i}}v+uv^{2^{2i}})x^{2^i}+C_2^{2^i}x=u^{2^{2i}}C_2+uC_2^{2^{i}}.
  \]
  Divide both sides by $u^{2^{2i}+2^i+1}$, then we have
  \begin{equation}\label{equ:solx2}
   \begin{split}
      & \left(\frac{v}{u}+\left(\frac{v}{u}\right)^{2^i}\right)\left(\frac{x}{u}\right)^{2^{2i}}+%
        \left(\frac{v}{u}+\left(\frac{v}{u}\right)^{2^{2i}}\right)\left(\frac{x}{u}\right)^{2^i}+%
        \left(\left(\frac{v}{u}\right)^{2^{i}}+\left(\frac{v}{u}\right)^{2^{2i}}\right)\frac{x}{u}\\
   =\ & \frac{v}{u}+\left(\frac{v}{u}\right)^{2^i}+\left(\frac{v}{u}\right)^{2^i}+\left(\frac{v}{u}\right)^{2^{2i}}.
   \end{split}
  \end{equation}
  Denote $w=\frac{v}{u}+\left(\frac{v}{u}\right)^{2^i}, z=\frac{x}{u}+\left(\frac{x}{u}\right)^{2^i}$. Then $w\neq 0$ since $u\neq 0, v\neq 0$ and $u\neq v$. The above equation is equivalent to
  \begin{equation}\label{equ:soluz}
   c(z+1)^{2^i}+c^{2^i}(z+1)=0.
  \end{equation}
  The solution of \eqref{equ:soluz} is $z=1$ or $z=w+1$ because $\gcd(i,k)=1$.\par

  If $z=1$, i.e., $\frac{x}{u}+\left(\frac{x}{u}\right)^{2^i}=1$. In this case, \eqref{equ:solx2} has no solution over $\GF{k}$. Otherwise, we have $\Tr(\frac{x}{u}+\left(\frac{x}{u}\right)^{2^i})=0\neq\Tr(1)$ since $k$ is odd.\par

  If $z=w+1$, i.e., $\frac{x}{u}+\left(\frac{x}{u}\right)^{2^i}=\frac{v}{u}+\left(\frac{v}{u}\right)^{2^i}+1$. In this case,  \eqref{equ:solx2} has no solution over $\GF{k}$ as well.\par

  This completes the proof.\QED
\end{prf}

\begin{rmk}
  We have also studied experimentially the bijective property of the closed butterfly structure with other $\alpha$. However, we could not find an $\alpha\neq 0,1$ such that $\Vb_{e}^{\alpha}$ is a permutation over $\GF{k}^2$. We conjecture that $\Vb_{e}^{\alpha}$ is a permutation over $\GF{k}^2$ if and only if $\alpha=1$.
\end{rmk}

\subsection{Other Cryptographic Properties}\label{sbs:crypp}

When $\alpha =1$ in the butterfly structure, the $\Hb_e^1$ is functionally equivalent to the 3-round Feistel structure constructed by Li and Wang \cite{FeistStrct}. They proved the differential spectrum is $\{0,4\}$ and the algebraic degree is $k$. Next we consider the nonlinearity. Firstly, we need the following results.

\begin{lem}[\cite{FeistStrct}]\label{lem:cdxy2}
  Suppose $k$ is an odd integer and $\gcd(i,k)=1$. Then for any $(c,d)\in\GF{k}^2$ with $(c,d)\neq (0,0)$, the following system of equations in $x,y$
  \begin{equation}\label{equ:cdxy1}
  \begin{cases}
    d^{2^i}x+(dx)^{2^{k-i}}+(c+d)^{2^i}y+((c+d)y)^{2^{k-i}}=0,\\
    c^{2^i}x+(cx)^{2^{k-i}}+d^{2^i}y+(dy)^{2^{k-i}}=0,
  \end{cases}
  \end{equation}
  has exactly 4 solutions over $\GF{k}^2$.
\end{lem}

 We call $(x,y)$ nonzero if $(x,y)\neq(0,0)$. Note that $(0,0)$ is always a solution of \eqref{equ:cdxy1}. So for any $(c,d)\in\GF{k}^2$ with $(c,d)\neq (0,0)$, \eqref{equ:cdxy1} has exactly 3 nonzero solutions over $\GF{k}^2$. One can easily verify that the three nonzero solutions are $(x,y),(y,x+y)$ and $(x+y,x)$ if $(x,y)$ is one of nonzero solutions of \eqref{equ:cdxy1}. \par

 Denote $S_{(a,b)}=\{(a,b),(b,a+b),(a+b,a)\}$. Obviously, any one element in $S_{(a,b)}$ determines completely the set, i.e., $S_{(a,b)}=S_{(b,a+b)}=S_{(a+b,a)}$. Furthermore for any $(a,b)\neq(0,0),(c,d)\neq(0,0)$, either we have $S_{(a,b)}=S_{(c,d)}$, or we have $S_{(a,b)}\cap S_{(c,d)}=\emptyset$. Put
 \[
  \mathcal{S}=\{S_{(a,b)}:(a,b)\in\GF{k}^2,(a,b)\neq (0,0)\},
 \]
 then obviously $\mathcal{S}$ is finite.

\begin{lem}\label{lem:cduv1}
  Suppose $k$ is an odd integer and $\gcd(i,k)=1$. Then for any $(c,d)\in\GF{k}^2$ with $(c,d)\neq (0,0)$, the following system of equations in variables $u$ and $v$
  \begin{equation}\label{equ:cduv1}
  \begin{cases}
    du^{2^i}+(du)^{2^{k-i}}+(c+d)v^{2^i}+((c+d)v)^{2^{k-i}}=0,\\
    (c+d)u^{2^i}+((c+d)u)^{2^{k-i}}+cv^{2^i}+(cv)^{2^{k-i}}=0
  \end{cases}
  \end{equation}
  has exactly 4 solutions over $\GF{k}^2$.
\end{lem}

\begin{prf}
  Firstly, we show that \eqref{equ:cduv1} has at most 4 solutions. We add the first equation to the second equation and obtain
  \begin{equation}\label{equ:cduv4}
    \begin{cases}
    du^{2^i}+(du)^{2^{k-i}}+(c+d)v^{2^i}+((c+d)v)^{2^{k-i}}=0,\\
    cu^{2^i}+(cu)^{2^{k-i}}+dv^{2^i}+(dv)^{2^{k-i}}=0,
  \end{cases}
  \end{equation}
  then raise both equations to the $2^i$th power, we have
  \begin{numcases}{}
    d^{2^i}u^{2^{2i}}+du+(c+d)^{2^i}v^{2^{2i}}+(c+d)v=0,\label{equ:cduv2}\\
    c^{2^i}u^{2^{2i}}+cu+d^{2^i}v^{2^{2i}}+dv=0.\label{equ:cduv3}
  \end{numcases}\par

  If $c=0$, then $d\neq 0$, \eqref{equ:cduv3} in $v$ has 2 solutions. For each solution $y$, \eqref{equ:cduv2} in $u$ has at moat 2 solutions. Hence, \eqref{equ:cduv1} has at most 4 solutions. The cases $d=0,c\neq 0$ and $c=d\neq 0$ is identical.\par

  Next, we suppose that $u\neq 0, v\neq 0$, and $u\neq v$. We add \eqref{equ:cduv2} multiplied by $c^{2^i}$ to \eqref{equ:cduv3} multiplied by $d^{2^i}$ to eliminate $u^{2^{2i}}$, then replace the $u$ into \eqref{equ:cduv2} and get $D_1v^{2^{4i}}+D_2v^{2^{2i}}+D_3v=0$, where $D_1=d^{2^i}\frac{(c^2+cd+d^2)^{2^{3i}}}{(c^{2^i}d+cd^{2^i})^{2^i}}\neq 0$. According to Corollary \ref{cor:4sol4}, this equation in $v$ has at most 4 solutions. Since the solution $u$ is uniquely determined by $v$, \eqref{equ:cduv1} has at most 4 solutions.\par

  Considering the following system of equations
  \begin{equation}\label{equ:uvxy1}
    \begin{cases}
      U^{2^i}X+(UX)^{2^{k-i}}+(U+V)^{2^i}Y+((U+V)Y)^{2^{k-i}}=0,\\
      V^{2^i}X+(VX)^{2^{k-i}}+U^{2^i}Y+(UY)^{2^{k-i}}=0,
    \end{cases}
  \end{equation}
  If we fix $(U,V)=(d,c)\neq(0,0)$, then \eqref{equ:uvxy1} with variables $X$ and $Y$ has exactly 3 nonzero solutions over $\GF{k}^2$ from Lemma \ref{lem:cdxy2}. W.l.o.g., suppose $(x,y),(y,x+y)$ and $(x+y,x)$ are the three nonzero solutions. If we fix $(U,V)=(c,c+d)$ or $(U,V)=(c+d,d)$, it is easy to verify that the three nonzero solutions of \eqref{equ:uvxy1} in $X$ and $Y$ are also $(x,y),(y,x+y)$ and $(x+y,x)$. \par

  Define a map\\
  \begin{center}
  \begin{tabular}{rrll}
    $\phi:$ & $\mathcal{S}$ & $\longrightarrow$  & $\mathcal{S}$ \\
            & $S_{(a,d)}$   & $\longmapsto$      & $S_{(x,y)}$, \\
  \end{tabular}
  \end{center}
  where $(x,y)$ is any nonzero solution of \eqref{equ:uvxy1} in variables $X$ and $Y$ with respect to coefficients $(U,V)=(a,b)$. This map is well-defined from above illustration. Then according to what we have showed, \eqref{equ:cduv4} has at most 4 solutions, which means $\phi$ is injective. But since $\mathcal{S}$ is finite, therefore $\phi$ is bijective. So for any $(d,c)\in\GF{k}^2,(d,c)\neq (0,0)$, there exists $(u,v)\neq (0,0)$ such that $\phi(S_{(u,v)})=S_{(d,c)}$, which is mean that if $(X,Y)=(d,c)$, then \eqref{equ:uvxy1} in variable $U$ and $V$ has exactly three nonzero solutions. \par

  We complete the proof.\QED
\end{prf}

The proof of nonlinearity is completely identical to the proof in Theorem \ref{thm:nolin}. From Lemma \ref{lem:cduv1}, we have $m=2$, the  $\mathcal{W}^2_F((a,b),(c,d))=0$, or $2^{2k+2}$. Therefore, $\Lambda_F=\{0,\pm 2^{k+1}\}$, and the nonlinearity $\mathcal{NL}(F)=2^{2k-1}-2^k$. \par

At the end of this section, we summarize the main results as follows.

\begin{thm}
  Suppose $k$ is an odd integer and $\gcd(i,k)=1$. For any $0\leq t\leq k-1$, let $\Hb_{e}^1$ and $\Vb_{e}^1$ be the open and closed $2k$-bit butterflies structure with exponent $e=(2^i+1)\times 2^t$. then\\
  (1) Both of $\Hb_{e}^1$ and $\Vb_{e}^1$ are permutation over $\GF{k}^2$.\\
  (2) The algebraic degree of $\Hb_{e}^1$ and $\Vb_{e}^1$ are equal to, respectively, $k$ and 2.\\
  (3) The differential uniformity of both $\Hb_{e}^1$ and $\Vb_{e}^1$ are equal to 4 and the differential spectrum are $\{0,4\}$.\\
  (4) The nonlinearity of both $\Hb_{e}^1$ and $\Vb_{e}^1$ are equal to $2^{2k-1}-2^k$, namely, optimal, and their Walsh spectrum are $\{0,\pm 2^{k+1}\}$.
\end{thm}

\section{Conclusion}\label{sec:concl}

In the present paper, we further study the butterfly structure and show that these structure always have very good cryptographic properties. Moveover, we prove the nonlinearity is optimal in the general case. The research of finding more classes of differentially 4-uniform permutations with highly nonlinearity and algebraic degree from other functions over subfields is very interesting and is worthy of a further investigation. The following questions is still open.\par
\textbf{Open Problems}:
Is there a tuple $k,e,\alpha$ where $k>3$ and $e$ are integers, and $\alpha$ is a finite field element such that $\Hb_e^{\alpha}$ operating on $\GF{k}^2$ is APN?


\section*{Appendix}

\noindent\textbf{The proof of $B_7\neq 0$ :}\par

Otherwise, we suppose that $B_7=0$. For simplify expressions, denote $\beta=\alpha^{2^i}$, we have
\[
(\beta c+\alpha d)B_6^{2^i}=(\alpha\beta d+c+d)B_4^{2^i}.
\]
Replace the above equation by the expressions $B_4$ and $B_6$, we have the following expression
\[
\begin{split}
    & (\beta c+\alpha d)^{2^{2i}}\left[(\beta c+\alpha d)(\beta c+\alpha d)^{2^i}+(\alpha\beta d+c+d)(\alpha\beta c+c+d)^{2^i}\right]\\
 =\ & (\alpha\beta c+c+d)^{2^{2i}}\left[(\beta c+\alpha d)(\alpha\beta d+c+d)^{2^i}+(\alpha\beta d+c+d)(\alpha c+\beta d)^{2^i}\right].
\end{split}
\]
Note that $\alpha\beta c+c+d\neq 0$ and $\beta c+\alpha d\neq 0$, we have
\begin{equation}\label{equ:abcd6}
(\beta c+\alpha d)(\alpha\beta d+c+d)^{2^i}+(\alpha\beta d+c+d)(\alpha c+\beta d)^{2^i}=\frac{(\beta c+\alpha d)^{2^{2i}}}{(\alpha\beta c+c+d)^{2^{2i}}}B_6.
\end{equation}\par

We also have $\alpha\beta d+c+d\neq 0$, otherwise, we must have $(\beta c+\alpha d)B_6^{2^i}=0$, which is impossible. From $B_5=0$, we get
\[
\begin{split}
     & (\alpha\beta d+c+d)(\beta c+\alpha d)(\alpha c+\beta d)^{2^i}(\beta c+\alpha d)^{2^i}\\
  =\ & (\alpha\beta d+c+d)(\beta c+\alpha d)(\alpha\beta c+c+d)^{2^i}(\alpha\beta d+c+d)^{2^i}.
\end{split}
\]
Replace $(\beta c+\alpha d)(\beta c+\alpha d)^{2^i}=B_6+(\alpha\beta c+c+d)^{2^i}(\alpha\beta d+c+d)$ into the above equation and obtain
\[
\begin{split}
      & B_6(\alpha\beta d+c+d)(\alpha c+\beta d)^{2^i}\\
  =\  & (\alpha\beta c+c+d)^{2^i}(\alpha\beta d+c+d)\left[(\beta c + \alpha d)(\alpha\beta d+c+d)^{2^i}+(\alpha\beta d+c+d)(\alpha c+\beta d)^{2^i}\right].
\end{split}
\]
According to \eqref{equ:abcd6} and $B_6\neq 0$, we deduce that
\[
(\alpha\beta d+c+d)(\alpha c+\beta d)^{2^i}=(\alpha\beta c+c+d)^{2^i}(\alpha\beta d+c+d)\frac{(\beta c+\alpha d)^{2^{2i}}}{(\alpha\beta c+c+d)^{2^{2i}}},
\]
which is equivalent to \
\[
(\alpha c+\beta d)(\alpha\beta c+c+d)^{2^i}=(\alpha\beta c+c+d)(\beta c+\alpha d)^{2^i},
\]
meaning $B_4=0$, a contradiction. Hence, we complete the proof.\QED

\end{document}